# Invariant quantities of a Mueller matrix under rotation and retarder transformations

**José J. Gil**
*Universidad de Zaragoza. Pedro Cerbuna 12, 50009 Zaragoza, Spain*
*ppgil@unizar.es*

**Abstract**

Mueller matrices are defined with respect to appropriate Cartesian reference frames for the representation of the states of polarization of the input and output electromagnetic beams. The polarimetric quantities that are invariant under rotations of the said reference frames about the respective directions of propagation (rotation transformations) provide particularly interesting physical information. Moreover, certain properties are also invariant with respect to the action of birefringent devices located at both sides of the medium under consideration (retarder transformations). The polarimetric properties that remain invariant under rotation and retarder transformations are calculated from any given Mueller matrix and are then analyzed and interpreted, providing significant parameterizations of Mueller matrices in terms of meaningful physical quantities.

## 1. Introduction

Polarimetry constitutes today a consolidated core of knowledge and technologies with more and more applications in a very wide range of scientific and industrial areas. Thus, any advance in the understanding and interpretation of the information provided by polarimetric measurements has a high potential impact in their exploitation.

The elements $m_{kl}$ of **M** contain, in an implicit manner, physical information beyond their mere role as coefficients of the corresponding linear transformation of the input Stokes vectors into the output ones. The physical quantities involved in a given Mueller matrix **M**, like the diattenuation, the polarizance, the depolarization index, the indices of polarimetric purity and some other interesting parameters can be identified and properly defined from $m_{kl}$. Therefore, it is worth to identify and interpret the complete sets of quantities that are invariant under transformations related to rotations of the input and output reference frames (*rotation transformations*) or to the serial combination of retarders with the medium represented by **M** (*retarder transformations*). With the term "retarder" we refer to general, or elliptic, retarders. All these analyses result in the parameterization of **M** in terms of quantities with specific physical meaning.

Beyond the main objective of representing the information contained in **M** by means of physically significant quantities, it is worth to mention that the results contained in this work allows to get appropriate interpretations of endoscopic (fiber) polarimetric experiments. Indeed, in this specific measurement configuration, the Mueller matrix **M** of the sample is obtained through one- or two-way optical fiber(s) that behave most generally as elliptic retarders, wherefrom the potential importance of the derived invariants from **M**.

## 2. Significant physical quantities involved in a Mueller matrix

This section is devoted to summarize some polarimetric quantities defined previously by a number of authors. Such quantities, together with some additional parameters defined in further sections, are necessary to formalize and interpret the invariance properties of Mueller matrices, which constitute the main aim of this work.

To simplify certain mathematical expressions, let us first recall the following block expression of a Mueller matrix [1],

$$\mathbf{M} = m_{00} \begin{pmatrix} 1 & \mathbf{D}^T \\ \mathbf{P} & \mathbf{m} \end{pmatrix},$$

$$\mathbf{D} \equiv \frac{1}{m_{00}}(m_{01}, m_{02}, m_{03})^T, \quad \mathbf{P} \equiv \frac{1}{m_{00}}(m_{10}, m_{20}, m_{30})^T, \quad \mathbf{m} \equiv \frac{1}{m_{00}} \begin{pmatrix} m_{11} & m_{12} & m_{13} \\ m_{21} & m_{22} & m_{23} \\ m_{31} & m_{32} & m_{33} \end{pmatrix}, \tag{1}$$

where the vectors $\mathbf{D}$ and $\mathbf{P}$ are called respectively the *diattenuation vector* and the *polarizance vector* of $\mathbf{M}$ [2]. The absolute values of these vectors are called *diattenuation*, $D \equiv |\mathbf{D}|$ and *polarizance*, $P \equiv |\mathbf{P}|$. The *average intensity coefficient* of $\mathbf{M}$ (i.e., transmittance or reflectance for unpolarized input states) is given by $m_{00}$. For some purposes it is useful to use the following normalized version of $\mathbf{M}$

$$\hat{\mathbf{M}} \equiv \mathbf{M}/m_{00} = \begin{pmatrix} 1 & \mathbf{D}^T \\ \mathbf{P} & \mathbf{m} \end{pmatrix}. \tag{2}$$

The *degree of polarimetric purity* of $\mathbf{M}$ is given by the *depolarization index* [3], which can be expressed as

$$P_\Delta = \sqrt{\left(D^2 + P^2 + 3P_S^2\right)/3}, \tag{3}$$

where $P_S$ is the *degree of spherical purity*, defined as [4]

$$P_S \equiv \frac{1}{\sqrt{3}} \sqrt{\sum_{k,l=1}^{3} m_{kl}^2}, \tag{4}$$

so that $P_S$ provides a measure of the contribution to $P_\Delta$ that is not directly related to the diattenuation-polarizance properties.

The value of $P_S$ is restricted to the range $0 \leq P_S \leq 1$. The maximum value $P_S = 1$ entails $P_\Delta = 1$ and corresponds to a pure retarder $\mathbf{M}_R$. The minimum value of $P_S$ compatible with total polarimetric purity of the system ($P_\Delta = 1$) is $P_S = 1/\sqrt{3}$. The value $P_S = 0$ is attained when all the elements of the submatrix $\mathbf{m}$ are zero ($m_{kl} = 0; k, l = 1, 2, 3$).

Mueller matrices associated with systems that do not depolarize any totally polarized input state (i.e., whose depolarization index satisfies $P_\Delta = 1$) are called *pure Mueller matrices* (also called *Mueller-Jones matrices*), while Mueller matrices satisfying $P_\Delta < 1$ are called *nonpure* or *depolarizing* Mueller matrices.

Both polarizance *P* and diattenuation *D* of a Mueller matrix have a dual nature depending on the direction of propagation of light (forward or reverse) [5]; *D* is both the diattenuation of **M** and the polarizance of the *reverse Mueller matrix* $\mathbf{M}^r \equiv \mathrm{diag}(1,1,-1,1)\,\mathbf{M}^T\,\mathrm{diag}(1,1,-1,1)$ [6,7] corresponding to the same interaction as **M** but interchanging the input and output directions. Despite *P* and *D* have respective specific physical meanings, for some purposes it is useful to consider the *degree of polarizance* $P_P$ [4]

$$P_P \equiv \sqrt{(P^2 + D^2)/2}, \tag{5}$$

which provides a measure of the joint contribution of *P* and *D* to polarimetric purity.

The value of $P_P$ is restricted to the range $0 \le P_P \le 1$, so that $P_P = 1$ corresponds to a pure polarizer. The value $P_P = 0$ corresponds to a *nonpolarizing Mueller matrix* **M** with zero diattenuation and zero polarizance.

Other quantities that will be useful for characterizing certain invariant properties of **M** are the *linear diattenuation* $D_L$ and the *circular diattenuation* $D_C$, defined as the respective *degree of linear polarization* and *degree of circular polarization* of the Stokes vector $\mathbf{s}_D \equiv (1, \mathbf{D}^T)^T$

$$D_L \equiv \sqrt{D_1^2 + D_2^2}, \quad D_C \equiv D_3, \tag{6}$$

where $D_i$ are the components of the diattenuation vector $\mathbf{D} \equiv (D_1, D_2, D_3)^T$.

When $D_C = 0$ the diattenuation affects exclusively the linear components $s_1$ and $s_2$ of the input Stokes vector $\mathbf{s} \equiv (s_0, s_1, s_2, s_3)^T$. Conversely, when $D_L = 0$ the diattenuation only affects the circular component $s_3$. Observe that $D_L$ is defined as a nonnegative parameter, while $D_C$ takes values in the range $-1 \le D_C \le 1$.

The *linear polarizance* $P_L$ and *circular polarizance* $P_C$ are defined as the respective *degree of linear polarization* and *degree of circular polarization* of the Stokes vector $\mathbf{s}_P \equiv \mathbf{M}\mathbf{s}_u$, image of the input unpolarized state $\mathbf{s}_u \equiv (1, \mathbf{0}^T)^T$

$$P_L \equiv \sqrt{P_1^2 + P_2^2}, \quad P_C \equiv P_3, \tag{7}$$

where $P_i$ are the components of the polarizance vector $\mathbf{P} \equiv (P_1, P_2, P_3)^T$

When $P_C = 0$ the polarizance affects exclusively the linear components of the Stokes vector $\mathbf{s}_P$. Conversely, when $P_L = 0$ the polarizance only affects the circular component of $\mathbf{s}_P$. As with the components of diattenuation, $P_L \ge 0$ while $-1 \le P_C \le 1$.

Moreover, it is also worth considering the singular value decomposition of the 3x3 submatrix **m** of **M**

$$\mathbf{m} = \mathbf{m}_{RO}\,\mathbf{m}_A\,\mathbf{m}_{RI}, \tag{8.a}$$

$\mathbf{m}_{RI}$ and $\mathbf{m}_{RO}$ being proper orthogonal matrices (i.e. $\det \mathbf{m}_{RI} = \det \mathbf{m}_{RO} = +1$), and $\mathbf{m}_A$ being the diagonal matrix whose entries are defined as follows from the singular values of $\mathbf{m}$, taken in decreasing order,

$$\mathbf{m}_A \equiv \operatorname{diag}(a_1, a_2, \varepsilon a_3), \quad a_1 \geq a_2 \geq a_3 \geq 0, \quad \varepsilon \equiv (\det \mathbf{M})/|\det \mathbf{M}|, \tag{8.b}$$

so that the *arrow decomposition* of $\mathbf{M}$ is defined as [8]

$$\mathbf{M} = \mathbf{M}_{RO} \mathbf{M}_A \mathbf{M}_{RI}, \tag{9}$$

where

$$\mathbf{M}_{RO} \equiv \begin{pmatrix} 1 & \mathbf{0}^T \\ \mathbf{0} & \mathbf{m}_{RO} \end{pmatrix}, \quad \mathbf{M}_{RI} \equiv \begin{pmatrix} 1 & \mathbf{0}^T \\ \mathbf{0} & \mathbf{m}_{RI} \end{pmatrix}, \tag{10}$$

represent Mueller matrices of respective retarders (in general, elliptic), while the *arrow form* $\mathbf{M}_A$ of $\mathbf{M}$ is given by

$$\mathbf{M}_A \equiv m_{00} \begin{pmatrix} 1 & \mathbf{D}_A^T \\ \mathbf{P}_A & \mathbf{m}_A \end{pmatrix} = \mathbf{M}_{RO}^T \mathbf{M} \mathbf{M}_{RI}^T = m_{00} \begin{pmatrix} 1 & \mathbf{D}^T \mathbf{m}_{RI}^T \\ \mathbf{m}_{RO}^T \mathbf{P} & \mathbf{m}_{RO}^T \mathbf{m} \mathbf{m}_{RI}^T \end{pmatrix}. \tag{11}$$

Note that $D_A = D$, $P_A = P$, and

$$D_A = D, \quad P_A = P, \quad P_S(\mathbf{M}) = P_S(\mathbf{M}_A) = \sqrt{(a_1^2 + a_2^2 + a_3^2)/3}. \tag{12}$$

The feasible region for the values of the *components of purity* $P_P$ and $P_S$ together with a detailed case analysis can be found in Ref. [4].

To complete this summary of relevant polarimetric quantities derivable from a given Mueller matrix $\mathbf{M}$, and whose invariance under certain transformations will be considered in subsequent sections, let us recall that $\mathbf{M}$ has a biunivocal relation with its associated covariance matrix $\mathbf{H}$, which can be submitted to the *spectral* or *Cloude's decomposition* [9]

$$\mathbf{H} = (\operatorname{tr} \mathbf{H}) \sum_{i=1}^{4} \hat{\lambda}_i \mathbf{H}_{Ji}, \quad \left( \mathbf{H}_{Ji} \equiv \mathbf{u}_i \otimes \mathbf{u}_i^\dagger, \quad \hat{\lambda}_i \equiv \lambda_i/(\operatorname{tr} \mathbf{H}) = \lambda_i/m_{00} \right), \tag{13}$$

where $\otimes$ stands for the Kronecker product, the superscript "$\dagger$" indicates conjugate transposed, $\lambda_i$ are the ordered eigenvalues of $\mathbf{H}$ $(\lambda_1 \geq \lambda_2 \geq \lambda_3 \geq \lambda_4 \geq 0)$, and $\mathbf{u}_i$ are the corresponding eigenvectors of $\mathbf{H}$. Note also that the subscript *J* of the addends $\mathbf{H}_{Ji}$ has been used to stress that they have only one nonzero eigenvalue and thus are associated with nondepolarizing (or pure) Mueller matrices $\mathbf{M}_{Ji}$. Thus, the *spectral decomposition* of $\mathbf{M}$ is formulated as [9]

$$\mathbf{M} = m_{00} \sum_{i=1}^{4} \hat{\lambda}_i \hat{\mathbf{M}}_{Ji}, \tag{14}$$

where $\hat{\mathbf{M}}_{Ji}$ are the pure Mueller matrices associated with $\mathbf{H}_{Ji}$. Note that, as pointed out in Refs. [5,10] the spectral decomposition, as well as other possible parallel decompositions of $\mathbf{M}$, to have physical consistency, requires to be expressed as a convex sum (i.e. the coefficients are nonnegative and sum to one) where the addends and $\mathbf{M}$ must have the same *mean intensity factor* (or mean transmittance) $m_{00}$.

Any pure Mueller matrix satisfies the equality $\mathrm{tr}(\mathbf{M}^T\mathbf{M}) = 4m_{00}$ [11], which is equivalent to $\lambda_2 = \lambda_3 = \lambda_4 = 0$ (and hence is equivalent to $P_\Delta = 1$). Moreover, the Mueller matrix of an ideal depolarizer is of the form $\mathbf{M}_{\Delta 0} \equiv \mathrm{diag}(m_{00}, 0, 0, 0)$, which is equivalent to $\lambda_1 = \lambda_2 = \lambda_3 = \lambda_4 = m_{00}/4$ (and hence is equivalent to $P_\Delta = 0$). In general, depolarizing Mueller matrices correspond to intermediate cases with $0 < P_\Delta < 1$ where the relative weights $\hat{\lambda}_i$ of the components in the spectral decomposition take arbitrary values (providing they satisfy the condition $\sum_{i=1}^{4} \hat{\lambda}_i = 1$, derived from their very definition). These considerations reflect the fact that depolarization properties are intrinsically related to the structure of the normalized eigenvalues $\hat{\lambda}_i$ of **H**. Therefore, a complete description of the polarimetric purity (i.e. polarimetric lack of randomness) of a medium requires considering three independent parameters derived from $\hat{\lambda}_i$.

At this point it should be noted that the depolarization index $P_\Delta$, as well as other alternative overall measures of the closeness of **M** to a pure Mueller matrix, like the Lorentz depolarization indices $L_1$ and $L_2$ [12], and the polarization entropy $S$ [13], do not provide complete information to deduce $\hat{\lambda}_i$. Moreover, $P$, $D$, $P_S$ give specific knowledge of the sources of purity, but are not sufficient to derive $\hat{\lambda}_i$.

An appropriate set of three invariant and dimensionless *indices of polarimetric purity* (IPP) providing complete information of the structure of polarimetric purity of **M** in terms of $\hat{\lambda}_i$ is defined as [14]

$$\mathcal{P}_1 \equiv \hat{\lambda}_1 - \hat{\lambda}_2, \quad \mathcal{P}_2 \equiv \hat{\lambda}_1 + \hat{\lambda}_2 - 2\hat{\lambda}_3, \quad \mathcal{P}_3 \equiv \hat{\lambda}_1 + \hat{\lambda}_2 + \hat{\lambda}_3 - 3\hat{\lambda}_4, \qquad (15)$$

(recall that the four nonnegative eigenvalues $\lambda_i$ of **H** have been taken ordered in decreasing magnitude $\lambda_1 \geq \lambda_2 \geq \lambda_3 \geq \lambda_4$ and that that this choice of ordering must be preserved to maintain the mathematical and physical meaning of the IPP).

As expected, the depolarization index $P_\Delta$ is not independent of the IPP, and can be calculated as follows

$$P_\Delta^2 = \frac{1}{3}\left(2\mathcal{P}_1^2 + \frac{2}{3}\mathcal{P}_2^2 + \frac{1}{3}\mathcal{P}_3^2\right), \qquad (16)$$

while the IPP are restricted by the following nested inequalities [14]

$$0 \leq \mathcal{P}_1 \leq \mathcal{P}_2 \leq \mathcal{P}_3 \leq 1. \qquad (17)$$

Thus, if $\mathcal{P}_1 = 1$, then $\mathcal{P}_2 = \mathcal{P}_3 = P_\Delta = 1$ (pure system). Moreover, if $\mathcal{P}_3 = 0$, then $\mathcal{P}_1 = \mathcal{P}_2 = P_\Delta = 0$ (ideal depolarizer). In general, the IPP provide complete information about polarimetric purity in terms of the weights of the spectral components of **M**, and constitute a representation complementary to that of $D$, $P$, and $P_S$ because the IPP are insensitive to the nature of the medium with respect to diattenuation, polarizance and retardance. In particular, the IPP can be interpreted physically from the so-called *characteristic* (or *trivial*) decomposition of **M** [5,15],

$$\mathbf{M} = \mathcal{P}_1 m_{00} \hat{\mathbf{M}}_J(\hat{\mathbf{H}}_J) + (\mathcal{P}_2 - \mathcal{P}_1) m_{00} \hat{\mathbf{M}}_2(\hat{\mathbf{H}}_2) + (\mathcal{P}_3 - \mathcal{P}_2) m_{00} \hat{\mathbf{M}}_3(\hat{\mathbf{H}}_3) + (1 - \mathcal{P}_3) m_{00} \hat{\mathbf{M}}_{\Delta 0}(\hat{\mathbf{H}}_{\Delta 0}), \qquad (18)$$

so that,

- $\mathbf{M}_J \equiv m_{00}\hat{\mathbf{M}}_J$ is the *characteristic pure component*, whose associated covariance matrix $\mathbf{H}_J \equiv m_{00}\left(\mathbf{u}_1 \otimes \mathbf{u}_1^\dagger\right)$ is defined from the eigenvector $\mathbf{u}_1$ with the largest eigenvalue of the covariance matrix $\mathbf{H}$ associated with $\mathbf{M}$. The relative weight of $\mathbf{M}_J$ with respect to the complete $\mathbf{M}$ is given by the first index of polarimetric purity $\mathcal{P}_1$ of $\mathbf{M}$.

- $\mathbf{M}_2 \equiv m_{00}\hat{\mathbf{M}}_2$, represents a *2D depolarizer*, whose associated covariance matrix $\mathbf{H}_2 \equiv \frac{1}{2}m_{00}\sum_{i=1}^{2}\mathbf{u}_i \otimes \mathbf{u}_i^\dagger$ has two equal (nonzero) eigenvalues and two zero eigenvalues, in such a manner that $\mathbf{M}_2$ is constituted by an equiprobable mixture of two pure components (namely, the first two spectral components). The relative weight of $\mathbf{M}_2$ with respect to the whole $\mathbf{M}$ is given by the difference $\mathcal{P}_2 - \mathcal{P}_1$ between the first two IPP.

- $\mathbf{M}_3 \equiv m_{00}\hat{\mathbf{M}}_3$, represents a *3D depolarizer*, whose associated covariance matrix $\mathbf{H}_3 \equiv \frac{1}{3}m_{00}\sum_{i=1}^{3}\mathbf{u}_i \otimes \mathbf{u}_i^\dagger$ has three equal (nonzero) eigenvalues and a zero eigenvalue, in such a manner that $\mathbf{M}_3$ is constituted by an equiprobable mixture of three pure components (namely, the first three spectral components). The relative weight of $\mathbf{M}_3$ with respect to the whole $\mathbf{M}$ is given by the difference $\mathcal{P}_3 - \mathcal{P}_2$.

- $\mathbf{M}_{\Delta 0} \equiv m_{00}\,\mathrm{diag}(1,0,0,0)$ represents an *ideal depolarizer* (or *4D depolarizer*), whose associated covariance matrix $\mathbf{H}_{\Delta 0} \equiv \frac{1}{4}m_{00}\mathbf{I}$ ($\mathbf{I}$ being the identity matrix) has four equal (nonzero) eigenvalues, in such a manner that $\mathbf{M}_{\Delta 0}$ is constituted by an equiprobable mixture of four pure components (namely, the four spectral components). The relative weight of $\mathbf{M}_{\Delta 0}$ with respect to the whole $\mathbf{M}$ is given by the difference $1 - \mathcal{P}_3$.

Note that, in accordance with the above considerations, the sole knowledge of $\mathcal{P}_1$, $\mathcal{P}_2$ and $\mathcal{P}_3$ does not imply necessarily the knowledge of $P$, $D$ and $P_S$; nevertheless, the set of five quantities $(\mathcal{P}_1, \mathcal{P}_2, \mathcal{P}_3, P, D)$ is sufficient to calculate $P_\Delta$ and $P_S$ [15].

## 3. Changes of reference frame and rotated Mueller matrices

In general, when dealing with a polarimetric interaction $\mathbf{s}' = \mathbf{M}\mathbf{s}$, represented through the Stokes-Mueller formalism, the input and output states are referred to with respect to local reference frames which include the respective directions of propagation as the reference axes $Z$ and $Z'$ and are usually represented, as in Fig. 1, through a model where the input and output directions of propagation have been disposed along a common $Z$ axis.

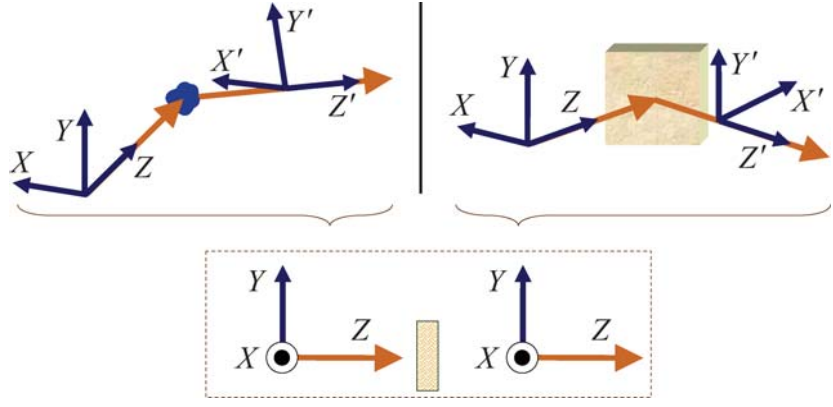

Fig. 1. Local reference frames for input and output polarization states. Examples for scattering and reflection.

When considering the relation $\mathbf{Ms} = \mathbf{s}'$, but referred to a coordinate system $X'Y'$ rotated at a counterclockwise angle $\theta$ with respect to the original $XY$, it is expressed as $\mathbf{M}_{X'Y'}\mathbf{s}_{X'Y'} = \mathbf{s}'_{X'Y'}$ where

$$\mathbf{M}_{X'Y'} = \mathbf{M}_G(\theta)\mathbf{M}\,\mathbf{M}_G(-\theta), \tag{19}$$

$\mathbf{M}_G(\theta)$ being the rotation matrix

$$\mathbf{M}_G(\theta) \equiv \begin{pmatrix} 1 & 0 & 0 & 0 \\ 0 & \cos 2\theta & \sin 2\theta & 0 \\ 0 & -\sin 2\theta & \cos 2\theta & 0 \\ 0 & 0 & 0 & 1 \end{pmatrix}. \tag{20}$$

## 4. Dual-retarder transformation

Given a Mueller matrix $\mathbf{M}$ and an arbitrary pair of orthogonal Mueller matrices $(\mathbf{M}_{R1}, \mathbf{M}_{R2})$, the matrix $\mathbf{M}'$ obtained through the *dual-retarder transformation*

$$\mathbf{M}' = \mathbf{M}_{R2}\mathbf{M}\,\mathbf{M}_{R1} = \begin{pmatrix} 1 & \mathbf{0}^T \\ \mathbf{0} & \mathbf{m}_{R2} \end{pmatrix} m_{00} \begin{pmatrix} 1 & \mathbf{D}^T \\ \mathbf{P} & \mathbf{m} \end{pmatrix} \begin{pmatrix} 1 & \mathbf{0}^T \\ \mathbf{0} & \mathbf{m}_{R1} \end{pmatrix} = m_{00} \begin{pmatrix} 1 & \mathbf{D}^T \mathbf{m}_{R1} \\ \mathbf{m}_{R2}\mathbf{P} & \mathbf{m}_{R2}\mathbf{m}\,\mathbf{m}_{R1} \end{pmatrix}, \tag{21}$$

is called *invariant-equivalent* to $\mathbf{M}$. Observe that the following parameters (namely, the average intensity factor $m_{00}$, the diattenuation $D$, the polarizance $P$, the degree of spherical purity $P_S$, the scalar quantity $\mathbf{P}^T\mathbf{m}\mathbf{D}$ and $\det\mathbf{M}$) remain unchanged under the dual retarder transformation

$$\begin{aligned}
&(\mathbf{M}')_{00} = m_{00} \\
&|\mathbf{D}'| = \sqrt{\mathbf{D}'^T\mathbf{D}'} = \sqrt{\mathbf{D}^T\mathbf{m}_{R1}\mathbf{m}_{R1}^T\mathbf{D}} = \sqrt{\mathbf{D}^T\mathbf{D}} = |\mathbf{D}| \\
&|\mathbf{P}'| = \sqrt{\mathbf{P}'^T\mathbf{P}'} = \sqrt{\mathbf{P}^T\mathbf{m}_{R2}\mathbf{m}_{R2}^T\mathbf{P}} = \sqrt{\mathbf{P}^T\mathbf{P}} = |\mathbf{P}| \\
&P'_S = \frac{1}{\sqrt{3}}\|\mathbf{m}'\|_2 = \frac{1}{\sqrt{3}}\sqrt{\operatorname{tr}(\mathbf{m}'^T\mathbf{m}')} = \frac{1}{\sqrt{3}}\sqrt{\operatorname{tr}(\mathbf{m}_{R1}^T\mathbf{m}^T\mathbf{m}_{R2}^T\mathbf{m}_{R2}\mathbf{m}\mathbf{m}_{R1})} = \frac{1}{\sqrt{3}}\sqrt{\operatorname{tr}(\mathbf{m}^T\mathbf{m})} = P_S \\
&\mathbf{P}'^T\mathbf{m}'\mathbf{D}' = \mathbf{P}^T\mathbf{m}_{R2}^T\mathbf{m}_{R2}\mathbf{m}\,\mathbf{m}_{R1}\mathbf{m}_{R1}^T\mathbf{D} = \mathbf{P}^T\mathbf{m}\mathbf{D} \\
&\det\mathbf{M}' = \det\mathbf{M}_{R2}\det\mathbf{M}\det\mathbf{M}_{R1} = \det\mathbf{M}
\end{aligned} \tag{22}$$

Obviously, $P_P$, $P_\Delta$ and other quantities derived from the previous ones are also invariant with respect to dual retarder transformations. In addition, it is straightforward to prove that the IPP as well as the singular values $(a_1, a_2, a_3)$ of **m** are also preserved.

Since, in general, a Mueller matrix **M** involves up to sixteen independent parameters (for instance its sixteen elements) and since $\mathbf{M}_{R1}$ and $\mathbf{M}_{R2}$ depend respectively up to three parameters, then the number of independent quantities that remain invariant under a dual retarder transformation cannot exceed from ten.

By taking into account Eq. (12) we see that $P_S$ is entirely determined by the set $(a_1, a_2, a_3)$, and, in addition, $P_\Delta$ can be obtained from $D$, $P$ and $P_S$, whereas one of the IPP (say, $\mathcal{P}_3$) can be obtained from $P_\Delta$ and the two remaining IPP.

In summary, a set of ten independent parameters remaining invariant under dual retarder transformations is given by the following quantities

$$m_{00},\ P,\ D,\ a_1,\ a_2,\ a_3,\ \mathcal{P}_1,\ \mathcal{P}_2,\ \mathbf{P}^T\mathbf{m}\mathbf{D},\ \det\mathbf{M}. \tag{23}$$

A number of invariant parameters can be obtained from the above set like, for instance,

$$P_P,\ P_S,\ P_\Delta,\ \mathcal{P}_3. \tag{24}$$

Since orthogonal Mueller matrices represent pure retarders, the transformation (21) can be physically realized by sandwiching the medium represented by **M** by the respective retarders. Therefore, by recalling that the product of orthogonal matrices is orthogonal, we conclude that any serial combination of a Mueller matrix **M** and an arbitrary number of retarders preserves the values of the above-mentioned set of quantities, which can be considered as *intrinsic* of **M**.

Concerning the *normal form* of **M** [16-18],

$$\mathbf{M} = \mathbf{M}_{J2}\mathbf{M}_\Delta\mathbf{M}_{J1}, \tag{25}$$

where $\mathbf{M}_{J1}$ and $\mathbf{M}_{J2}$ are pure Mueller matrices, and $\mathbf{M}_\Delta$ is the type-I or type-II [19,20] canonical depolarizer [21], note that, regardless of whether **M** is type-I or type-II, the central depolarizer $\mathbf{M}_\Delta$ is preserved under dual retarder transformations

$$\mathbf{M}' = \mathbf{M}_{R2}\mathbf{M}\mathbf{M}_{R1} = \mathbf{M}_{R2}(\mathbf{M}_{J2}\mathbf{M}_\Delta\mathbf{M}_{J1})\mathbf{M}_{R1} = (\mathbf{M}_{R2}\mathbf{M}_{J2})\mathbf{M}_\Delta(\mathbf{M}_{J1}\mathbf{M}_{R1}) = \mathbf{M}'_{J2}\mathbf{M}_\Delta\mathbf{M}'_{J1}. \tag{26}$$

Therefore, the eigenvalues $\rho_i$ of the N-matrix $\mathbf{N} = \mathbf{G}\mathbf{M}^T\mathbf{G}\mathbf{M}$, **G** being the Minkowski metric $\mathbf{G} \equiv \text{diag}(1,-1,-1,-1)$, from whose square roots $\sqrt{\rho_i}$ are defined the elements of $\mathbf{M}_\Delta$, also remain unchanged. Consequently, the first and second Lorentz depolarization indices $L_1(\mathbf{M})$ and $L_2(\mathbf{M})$ [12], are also invariant under dual-retarder transformations.

## 5. Single-retarder transformation

Let us now consider the *single-retarder transformation* of **M**, defined as an orthogonal similarity transformation

$$\mathbf{M}' = \mathbf{M}_R\mathbf{M}\mathbf{M}_R^T = \begin{pmatrix} 1 & \mathbf{0}^T \\ \mathbf{0} & \mathbf{m}_R \end{pmatrix} m_{00} \begin{pmatrix} 1 & \mathbf{D}^T \\ \mathbf{P} & \mathbf{m} \end{pmatrix} \begin{pmatrix} 1 & \mathbf{0}^T \\ \mathbf{0} & \mathbf{m}_R^T \end{pmatrix} = m_{00} \begin{pmatrix} 1 & \mathbf{D}^T\mathbf{m}_R^T \\ \mathbf{m}_R\mathbf{P} & \mathbf{m}_R\mathbf{m}\mathbf{m}_R^T \end{pmatrix}, \tag{27}$$

which can physically be performed by sandwiching the medium represented by **M** by two identical retarders whose fast eigenstates are mutually orthogonal (i.e., represented by antipodal points on the Poincaré sphere).

In this case, in addition to relations (22), the following equalities are satisfied

$$\mathbf{P}'^T \mathbf{D}' = \mathbf{P}^T \mathbf{m}_R^T \mathbf{m}_R \mathbf{D} = \mathbf{P}^T \mathbf{D}$$

$$\mathbf{P}'^T \mathbf{m}'^T \mathbf{D}' = \mathbf{P}^T \mathbf{m}_R^T \mathbf{m}_R \mathbf{m}^T \mathbf{m}_R^T \mathbf{m}_R \mathbf{D} = \mathbf{P}^T \mathbf{m}^T \mathbf{D} \quad (28)$$

$$\mathrm{tr}\,\mathbf{M}' = \mathrm{tr}\,\mathbf{M}$$

Since $\mathbf{M}_R$ depends on up to three parameters, thirteen is the maximum number of independent quantities that remain invariant under a single-retarder transformation, as occurs with the following set

$$m_{00},\ P,\ D,\ a_1,\ a_2,\ a_3,\ \mathcal{P}_1,\ \mathcal{P}_2,\ \mathbf{P}^T \mathbf{m} \mathbf{D},\ \mathbf{P}^T \mathbf{m}^T \mathbf{D},\ \mathbf{P}^T \mathbf{D},\ \mathrm{tr}\,\mathbf{M},\ \det\mathbf{M}. \quad (29)$$

## 6. Dual-rotation transformation

Rotation matrices of the form (20) have the peculiarity that also can be associated to circular retarders (with retardance $\Delta = 2\theta$). Consequently only a particular subset of the dual-retarder transformations can be performed through physical rotations of the input and output laboratory reference frames around the respective axes $Z$ and $Z'$ defining the input and output directions of propagation of the interacting electromagnetic wave (Fig. 2).

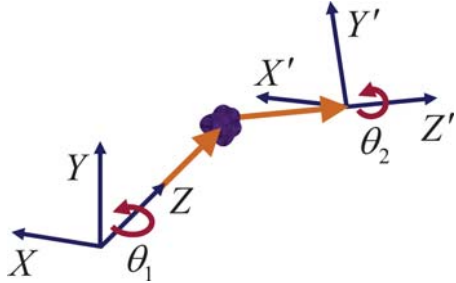

Figure 2. The *dual rotation transformation* consists of respective rotations $\theta_1$ and $\theta_2$ of the input and output reference frames around the respective propagation axes $Z$ and $Z'$.

This *dual-rotation transformation* corresponds to the case where the orthogonal Mueller matrices $\mathbf{M}_{R2}$ and $\mathbf{M}_{R1}$ in Eq. (21) have the form of respective rotation matrices $\mathbf{M}_{R2} = \mathbf{M}_G(\theta_2)$, $\mathbf{M}_{R1} = \mathbf{M}_G(\theta_1)$ [see Eq. (20)].

By taking advantage of the vectorial partitioned notation

$$\mathbf{M} \equiv m_{00} \begin{pmatrix} 1 & D_1 & D_2 & D_3 \\ P_1 & k_1 & r_3 & -r_2 \\ P_2 & q_3 & k_2 & r_1 \\ P_3 & -q_2 & q_1 & k_3 \end{pmatrix}, \quad (30)$$

in terms of the five vectors

$$\mathbf{D} \equiv (D_1, D_2, D_3)^T, \quad \mathbf{P} \equiv (P_1, P_2, P_3)^T,$$
$$\mathbf{k} \equiv (k_1, k_2, k_3)^T, \quad \mathbf{r} \equiv (r_1, r_2, r_3)^T, \quad \mathbf{q} \equiv (q_1, q_2, q_3)^T, \quad (31)$$

(all of them with absolute value less than one, and thus representable in the Poincaré sphere), the transformed Mueller matrix $\mathbf{M}'$ is given by

$$\mathbf{M}' = \mathbf{M}_{G2} \mathbf{M} \mathbf{M}_{G1} = \begin{pmatrix} 1 & \mathbf{0}^T \\ \mathbf{0} & \mathbf{m}_{G2} \end{pmatrix} m_{00} \begin{pmatrix} 1 & \mathbf{D}^T \\ \mathbf{P} & \mathbf{m} \end{pmatrix} \begin{pmatrix} 1 & \mathbf{0}^T \\ \mathbf{0} & \mathbf{m}_{G1} \end{pmatrix} = m_{00} \begin{pmatrix} 1 & \mathbf{D}^T \mathbf{m}_{G1} \\ \mathbf{m}_{G2} \mathbf{P} & \mathbf{m}_{G2} \mathbf{m} \mathbf{m}_{G1} \end{pmatrix}$$

$$= m_{00} \begin{pmatrix} 1 & c_1 D_1 + s_1 D_2 & -s_1 D_1 + c_1 D_2 & D_3 \\ c_2 P_1 + s_2 P_2 & c_2 (c_1 k_1 + s_1 r_3) + s_2 (c_1 q_3 + s_1 k_2) & c_2 (-s_1 k_1 + c_1 r_3) + s_2 (-s_1 q_3 + c_1 k_2) & -c_2 r_2 + s_2 r_1 \\ -s_2 P_1 + c_2 P_2 & -s_2 (c_1 k_1 + s_1 r_3) + c_2 (c_1 q_3 + s_1 k_2) & s_2 (s_1 k_1 - c_1 r_3) + c_2 (-s_1 q_3 + c_1 k_2) & s_2 r_2 + c_2 r_1 \\ P_3 & -c_1 q_2 + s_1 q_1 & s_1 q_2 + c_1 q_1 & k_3 \end{pmatrix} \quad (32)$$

where $s_i \equiv \sin 2\theta_i$ and $c_i \equiv \cos 2\theta_i$ ($i = 1, 2$).

Therefore, the following parameters (not all mutually independent), which include the set (23), remain invariant under dual-rotation transformations

$$m_{00}, D_L, D_C, D, P_L, P_C, P, P_P, r_L, r_C, r, q_L, q_C, q, a_1, a_2, a_3,$$
$$k_L^2 + r_C^2 + q_C^2, k_C, P_S, \mathcal{P}_1, \mathcal{P}_2, \mathcal{P}_3, P_\Delta, \mathbf{P}^T \mathbf{m} \mathbf{D}, \det \mathbf{M}, \quad (33)$$
$$\rho_0, \rho_1, \rho_2, \rho_3, L_1, L_2,$$

where the linear and circular components of $\mathbf{k}$, $\mathbf{r}$ and $\mathbf{q}$ have been defined as

$$k_L^2 \equiv k_1^2 + k_2^2, \quad r_L^2 \equiv r_1^2 + r_2^2, \quad q_L^2 \equiv q_1^2 + q_2^2,$$
$$k_C \equiv k_3, \quad r_C \equiv r_3, \quad q_C \equiv q_3. \quad (34)$$

Different sets of fourteen mutually independent parameters that remain invariant under dual retarder transformations can be chosen from (33), as for example

$$m_{00}, D_L, D_C, P_L, P_C, r_L, r_C, q_L, q_C, k_L^2 + r_C^2 + q_C^2, k_C, \mathcal{P}_1, \mathcal{P}_2, \det \mathbf{M}. \quad (35)$$

Therefore, a set of sixteen independent parameters providing all the information involved in $\mathbf{M}$ is completed by adding $\theta_1$ and $\theta_2$ to the above collection of invariant quantities.

## 7. Single-rotation transformation

A particular, but very common, type of dual rotation transformation occurs when $\theta_2 = -\theta_1$, which corresponds to a joint rotation of the input and output laboratory reference frames (Fig. 3), in which case the *single-rotation transformation* is formulated as

$$\mathbf{M}' = \mathbf{M}_G \mathbf{M} \mathbf{M}_G^T = \begin{pmatrix} 1 & \mathbf{0}^T \\ \mathbf{0} & \mathbf{m}_G \end{pmatrix} m_{00} \begin{pmatrix} 1 & \mathbf{D}^T \\ \mathbf{P} & \mathbf{m} \end{pmatrix} \begin{pmatrix} 1 & \mathbf{0}^T \\ \mathbf{0} & \mathbf{m}_G^T \end{pmatrix} = m_{00} \begin{pmatrix} 1 & \mathbf{D}^T \mathbf{m}_G^T \\ \mathbf{m}_G \mathbf{P} & \mathbf{m}_G \mathbf{m} \mathbf{m}_G^T \end{pmatrix}$$

$$= m_{00} \begin{pmatrix} 1 & cD_1 + sD_2 & -sD_1 + cD_2 & D_3 \\ cP_1 + sP_2 & c^2 k_1 + s^2 k_2 + sc(r_3 + q_3) & c^2 r_3 - s^2 q_3 + sc(k_2 - k_1) & -cr_2 + sr_1 \\ -sP_1 + cP_2 & -s^2 r_3 + c^2 q_3 + sc(k_2 - k_1) & s^2 k_1 + c^2 k_2 - sc(r_3 + q_3) & sr_2 + cr_1 \\ P_3 & -cq_2 + sq_1 & sq_2 + cq_1 & k_3 \end{pmatrix} \quad (36)$$

where $s \equiv \sin\theta$, $c \equiv \cos\theta$ $(\theta \equiv \theta_2 = -\theta_1)$.

Therefore, the following parameters (not all mutually independent), which include the sets (29) and (33) remain invariant under single-rotation transformations

$$m_{00}, D_L, D_C, D, P_L, P_C, P, P_P, r_L, r_C, r, q_L, q_C, q, k_L^2 + r_C^2 + q_C^2, r_C - q_C, k_C,$$
$$a_1, a_2, a_3, P_S, \mathcal{P}_1, \mathcal{P}_2, \mathcal{P}_3, P_\Delta, \mathbf{P}^T \mathbf{m} \mathbf{D}, \mathbf{P}^T \mathbf{m}^T \mathbf{D}, \mathbf{P}^T \mathbf{D}, \operatorname{tr}\mathbf{M}, \det\mathbf{M}, \quad (37)$$
$$\rho_0, \rho_1, \rho_2, \rho_3, L_1, L_2.$$

Several sets of fifteen mutually independent invariant parameters can be chosen from (37), as for example

$$m_{00}, D_L, D_C, P_L, P_C, r_L, r_C, q_L, q_C, k_L^2 + r_C^2 + q_C^2, k_C, \mathcal{P}_1, \mathcal{P}_2, \operatorname{tr}\mathbf{M}, \det\mathbf{M}, \quad (38)$$

or, alternatively, the following combination of the set of invariants (29) of the single-retarder transformation with the set (35) of invariants of the dual-rotation transformation

$$m_{00}, D_L, D_C, P_L, P_C, a_1, a_2, a_3, \mathbf{P}^T \mathbf{D}, \mathbf{P}^T \mathbf{m}^T \mathbf{D}, \mathbf{P}^T \mathbf{m} \mathbf{D}, \mathcal{P}_1, \mathcal{P}_2, \operatorname{tr}\mathbf{M}, \det\mathbf{M}. \quad (39)$$

Therefore, a set of sixteen independent parameters providing all the information involved in $\mathbf{M}$ is completed by adding $\theta$ to the above collections of fifteen invariants.

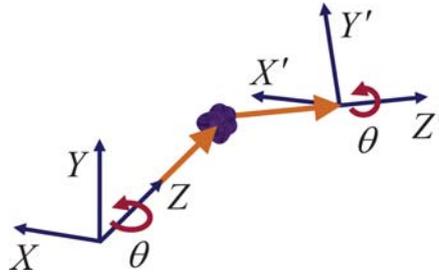

Figure 3. The *single rotation transformation* consists of respective rotations of a same angle $\theta$ of the input and output reference frames around the respective propagation axes $Z$ and $Z'$.

## Conclusion

At first sight, the Mueller matrix $\mathbf{M}$ seems to be the mere transfer matrix of the linear transformations of the Stokes vectors of the input states into those of the output states. Nevertheless, from the physical concept of Mueller matrix [9,22-23], the elements $m_{kl}$ of $\mathbf{M}$ contain, in an implicit manner, physical information beyond their role as coefficients of the corresponding linear transformation. In fact, complete sets of meaningful explicit quantities have been identified that are invariant under certain transformations related to rotations of the input and output reference frames (*rotation transformations*) or to the serial combination of retarders and the medium represented by **M** (*retarder transformations*). All these analyses result in the parameterization of **M** in terms of sixteen quantities with specific physical meaning. In fact, any Mueller matrix can be parameterized either by means of:

- the ten parameters (23) together with the 3+3 parameters characterizing the input and output retarders of the arrow decomposition (9) of **M**; or

- the thirteen parameters (29) together with the three parameters of a retarder; or

- the fourteen parameters (35) together with the two angles $\theta_1$ and $\theta_2$ corresponding to respective input and output rotations; or

- the fifteen parameters (39) together the angle $\theta$ corresponding to the input and output rotations.

In polarimetric arrangements where the input and output reference frames coincide (as is usually the case of polarimeters operating in transmission mode), the set (39) [or (38)] is independent of the orientation of the reference axes *XY* (*Z* being the direction of propagation of the probing electromagnetic beam). Since such orientation is arbitrary, the number of effective parameters to be analyzed after the measurement of **M** is reduced from 16 to 15. Further, unlike the elements $m_{ij}$ of **M**, the set (39) is constituted by quantities with physical significance like the linear and circular polarizance and diattenuation $P_L$, $P_C$, $D_L$ and $D_C$; the indices of polarimetric purity $\mathcal{P}_1$ and $\mathcal{P}_2$ (observe that $\mathcal{P}_3$ is also invariant). The physical meaning of $a_1, a_2, a_3$ is provided indirectly by the arrow decomposition of **M**. The determinant of **M** is indirectly related to $P$ and $D$, but it is not totally determined by them. Thus, $\det \mathbf{M}$ as well as other invariants that have been identified, like $\mathbf{P}^T \mathbf{D}$, $\mathbf{P}^T \mathbf{m}^T \mathbf{D}$, $\mathbf{P}^T \mathbf{m} \mathbf{D}$ and $\operatorname{tr} \mathbf{M}$ require an additional effort for identifying their specific physical interpretation. Observe, for instance that the invariant quantity $\mathbf{P}^T \mathbf{D}/(PD)$ gives a measure of the angle subtended between the polarizance and diattenuation vectors of **M**, that is, a measure of how close are the directions of the associated Stokes vectors in the Poincaré sphere.

When the interest is focused on the invariants under dual rotation, single retarder or dual retarder transformations, the number of effective parameters to be analyzed is reduced correspondingly and particular attention deserve the ten invariants of dual retarder transformations, which are also invariant in the other transformations considered and constitute the core of the invariant properties of a given Mueller matrix.

**Acknowledgement**

This research was supported by Ministerio de Economía y Competitividad and the European Union, grants FIS2011-22496 and FIS2014-58303-P, and by Gobierno de Aragón, group E99.

**References**

1. Z-F Xing, "On the deterministic and non-deterministic Mueller matrix" J. Mod Opt. **39**, 461-484 (1992).

2. S.-Y. Lu, R. A. Chipman, "Interpretation of Mueller matrices based on polar decomposition", J. Opt. Soc. Am. A **13**, 1106-1113 (1996).

3. J. J. Gil, E. Bernabéu, "Depolarization and polarization indices of an optical system" Opt. Acta **33**, 185-189 (1986).

4.  J. J. Gil, "Components of purity of a Mueller matrix" J. Opt. Soc. Am. A, 28, 1578-1585 (2011).

5.  J. J. Gil, "Polarimetric characterization of light and media" Eur. Phys. J. Appl. Phys. **40**, 1-47 (2007).

6.  Z. Sekera, "Scattering Matrices and Reciprocity Relationships for Various Representations of the State of Polarization," J. Opt. Soc. Am. **56**, 1732-1740 (1966).

7.  A. Schönhofer and H. -G. Kuball, "Symmetry properties of the Mueller matrix," Chem. Phys. **115**, pp. 159-167 (1987).

8.  J. J. Gil, "Transmittance constraints in serial decompositions of depolarizing Mueller matrices. The arrow form of a Mueller matrix," J. Opt. Soc. Am. A 30, 701-707 (2013).

9.  S.R. Cloude, "Group theory and polarization algebra", Optik **75**, 26-36 (1986).

10. J. J. Gil and I. San José, "Polarimetric subtraction of Mueller matrices," J Opt. Soc. Am. A **30**, 1078-1088 (2013).

11. J. J. Gil and E. Bernabéu, "A depolarization criterion in Mueller matrices," Opt. Acta **32**, 259–261 (1985).

12. R. Ossikovski, "Alternative depolarization criteria for Mueller matrices," J. Opt. Soc. Am. A **27**, 808-814 (2010).

13. S. R. Cloude, E. Pottier, "Concept of polarization entropy in optical scattering," Opt. Eng. **34**, 1599-1610 (1995).

14. I. San Jose, J. J. Gil, "Invariant indices of polarimetric purity. Generalized indices of purity for nxn covariance matrices" Opt. Commun. **284**, 38-47 (2011).

15. J. J. Gil, "Review on Mueller matrix algebra for the analysis of polarimetric measurements," J. Ap. Remote Sens. **8**, 081599-37 (2014).

16. R. Sridhar, and R. Simon, "Normal form for Mueller matrices in polarization optics," J. Mod. Opt. **41**, 1903-1915 (1994).

17. C. V. M. van der Mee, "An eigenvalue criterion for matrices transforming Stokes parameters," J. Math. Phys. **34**, 5072-5088 (1993).

18. R. Ossikovski, "Analysis of depolarizing Mueller matrices through a symmetric decomposition," J. Opt. Soc. Am. A **26**, 1109-1118 (2009).

19. A.V. Gopala Rao, K.S. Mallesh and Sudha, "On the algebraic characterization of a Mueller matrix in polarization optics. I. Identifying a Mueller matrix from its *N* matrix," J. Mod. Opt. **45**, 955-987 (1998).

20. A.V. Gopala Rao, K.S. Mallesh and Sudha, "On the algebraic characterization of a Mueller matrix in polarization optics. II. Necessary and sufficient conditions for Jones derived Mueller matrices," J. Mod. Opt. **45**, 989-999 (1998).

21. R. Ossikovski, "Canonical forms of depolarizing Mueller matrices", J. Opt. Soc. Am. A **27**, 123-130 (2010).

22. N. G. Parke III, *Matrix optics* Ph. D. thesis, M.I.T. 1948.

23. K. Kim, L. Mandel, E. Wolf "Relationship between Jones and Mueller matrices for random media" J. Opt. Soc. Am. A **4**, 433-437 (1987).

24. J. J. Gil, "Characteristic properties of Mueller matrices" J. Opt. Soc. Am. A **17**, 328-334 (2000).